\documentclass[prl,twocolumn,superscriptaddress,showpacs]{revtex4-1}
\usepackage{graphicx,amsmath,amssymb}
\usepackage{color}
\usepackage[normalem]{ulem}

\begin{document}

\title{A single-crystal neutron diffraction study on magnetic structure of the quasi-one-dimensional antiferromagnet SrCo$_{2}$V$_{2}$O$_{8}$}

\author{Juanjuan Liu}
\author{Jinchen Wang}
\author{Wei Luo}
\author{Jieming Sheng}
\affiliation{Department of Physics, Renmin University of China, Beijing
100872, China }
\author{Zhangzhen He}
\affiliation{State Key Laboratory of Structural Chemistry, Fujian Institute of Research on the Structure of Matter, Chinese Academy of Sciences, Fuzhou, Fujian 350002, P. R. China
}
\author{S.A. Danilkin }
\affiliation{Bragg Institute, ANSTO, Locked Bag 2001, Kirrawee DC NSW 2232, Australia
}

\author{Wei Bao$ ^1 $}
\email{wbao@ruc.edu.cn}

\date{\today}

\begin{abstract}
The magnetic structure of the spin-chain antiferromagnet SrCo$_{2}$V$_{2}$O$_{8}$ is determined by single-crystal neutron diffraction experiment. The system undergoes magnetic long range order below T$_{N}$ = 4.96 K. The moment of $2.16 \mu_{B}$ per Co at $1.6$ K in the screw chain running along the $c$ axis  alternates in the $c$-axis. The moments of neighboring screw chains are arranged antiferromagnetically along one in-plane axis and ferromagnetically along the other in-plane axis. This magnetic configuration breaks the 4-fold symmetry of the tetragonal crystal structure and leads to two equally populated magnetic twins with antiferromagnetic vector in the $a$ or $b$ axis. The very similar magnetic state to the isostructural
BaCo$_{2}$V$_{2}$O$_{8}$ warrants SrCo$_{2}$V$_{2}$O$_{8}$ another interesting half-integer spin-chain antiferromagnet for investigation on quantum antiferromagnetism.
\end{abstract}


\maketitle

Study on quasi-one-dimensional antiferromagnetic materials in recent decades has revealed 
interesting quantum magnetic effects which are absent in three-dimensional classical antiferromagnets.
While whether the spin is an integer or half-integer would have made no difference in the classical magnetic theory, 
drastically different physics phenomena have been predicted \cite{PhysRevLett.45.1358} and demonstrated between the
spin $S = 1$ systems \cite{JAP1.338595, PhysRevLett.56.371, PhysRevLett.69.3571, PhysRevB.50.9174, PhysRevB.84.054413} and the $S=\frac{1}{2}$ systems \cite{PhysRevB.52.13368, PhysRevLett.79.1750, giamarchi2008bose, PhysRevB.81.132401,PhysRevLett.87.017202}. 
In addition to chain compounds, magnetic moments forming a ladder constitute another important geometric system in investigation on quasi-one-dimensional quantum antiferromagnetism with the competition between the intra and inter-chain exchange interactions as an additional tuning parameter \cite{ PhysRevLett.101.137207, PhysRevLett.101.247202, PhysRevLett.111.106404, PhysRevLett.111.137205}.
More recently a series of quasi-one-dimensional spin systems built from a screw chain structure have been realized in
 transition metal vanadates  $AM_{2}$V$_{2}$O$_{8}$, where $A$ = Ba, Sr, or Pb and $M$ = Cu, Co, or Ni \cite{He2005486,he2005crystal,He2006458,He20071770}. 
Remarkably, a quantum phase transition to the Tomonaga-Luttinger liquid has been induced by an external magnetic field in BaCo$_{2}$V$_{2}$O$_{8}$  \cite{kimura2008novel,PhysRevB.76.224411,PhysRevLett.101.207201,PhysRevB.83.064421,PhysRevB.87.054408,PhysRevB.92.060408,PhysRevB.92.134416,PhysRevLett.114.017201}.
 
The magnetic structure of BaCo$_{2}$V$_{2}$O$_{8}$ at the zero magnetic field was first investigated by neutron powder diffraction experiments \cite{PhysRevB.83.064421} and completely determined by
neutron single-crystal diffraction experiments \cite{PhysRevB.87.054408}. The direct measurements of
the antiferromagnetic structure and its evolution in magnetic field by neutron diffraction experiments
have provided the valuable evidences to compare with theoretical prediction of the novel quantum phenomena in half-integer spin chains \cite{kimura2008novel,PhysRevB.76.224411,PhysRevLett.101.207201,PhysRevB.83.064421,PhysRevB.87.054408}.
For the isostructural SrCo$_{2}$V$_{2}$O$_{8}$, heat capacity and magnetic susceptibility measurements also indicate a magnetic transition at $T_N \sim 5$ K \cite{He2007667,Lejay2011128}.  An additional magnetic transition to a canted antiferromagnetic structure at $\sim 3$ K was reported in one of the studies \cite{He2007667}, which was different from the case for BaCo$_{2}$V$_{2}$O$_{8}$. A direct determination of the magnetic configuration by neutron diffraction is called for.
In this single-crystal neutron diffraction study of SrCo$_{2}$V$_{2}$O$_{8}$, we observed only one antiferromagnetic phase transition at $T_N = 4.96$ K. No anomaly in nuclear or magnetic Bragg peak intensity can be detected around 3 K. The magnetic structure of SrCo$_{2}$V$_{2}$O$_{8}$ is found to be the same as that of BaCo$_{2}$V$_{2}$O$_{8}$ reported previously \cite{PhysRevB.87.054408}. The magnetic moment 2.16(1) $\mu_{B}$ per Co$^{2+}$ ion in SrCo$_{2}$V$_{2}$O$_{8}$ is also very close to that of BaCo$_{2}$V$_{2}$O$_{8}$. Therefore, SrCo$_{2}$V$_{2}$O$_{8}$ can be another candidate of the vanadate series $AM_{2}$V$_{2}$O$_{8}$ for investigation on quasi-one-dimensional quantum antiferromagnetism.

\begin{figure}[bt]
\label{fig1}
\begin{center}
\includegraphics[width=80mm]{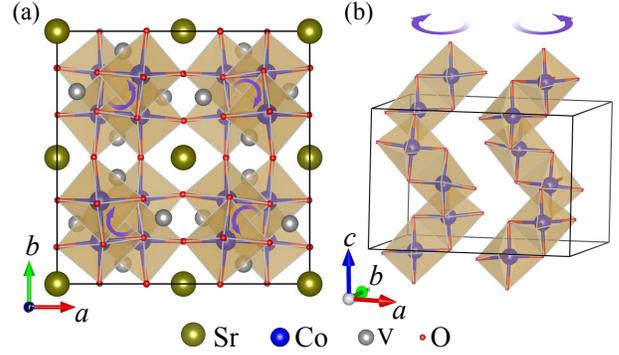}
\caption{(color online). Crystal structure of SrCo$_{2}$V$_{2}$O$_{8}$. (a) Projection to the (0,0,1) basal plane in the unit cell. (b) The magnetic chains are formed by edge sharing CoO$_{6}$ octahedra. Only two chains with  $0 \leq x \leq 1$ and $0 \leq y \leq 0.5$ are presented here. The arrows indicate the sense of rotation of the screw chains on increasing $z$.}
\end{center}
\end{figure}

The single crystal sample of SrCo$_{2}$V$_{2}$O$_{8}$ used in this study was grown by spontaneous nucleation method \cite{He2006458} with dimensions $\sim 7 \times 7 \times 7$ mm$^{3}$. 
Neutron scattering experiments were performed using the thermal triple-axis spectrometer Taipan at ANSTO \cite{danilkin_taipan_2007}. The horizontal collimations were open-$ 40' $-$ 40' $-open. Neutrons with incident energy $E$ = 14.7 meV were selected using the $ (0,0,2) $ reflection of a pyrolytic graphite (PG) monochromator. PG filters were used to remove higher-order neutrons from the neutron beam. The temperature of the sample was regulated by the ILL Orange Cryostat in a temperature range from 1.5 K to 300 K. 

SrCo$_{2}$V$_{2}$O$_{8}$ crystallizes in the tetragonal SrNi$_{2}$(AsO$_{4}$)$_{2}$ structure (space group No.\ 110, $I4_{1}cd$) \cite{CrystalStructure}, with lattice parameters $a = 12.267(1) \mathring{A}$ and $c = 8.424(1) \mathring{A}$ at room temperature. In a structural unit cell, the magnetic ions Co$^{2+}$ in the edge-shared CoO$_{6}$ octahedra form screw chains along the $c$-axis (Fig.~1\ref{fig1}). The screw chains are well separated by non-magnetic VO$_{4}$ tetrahedra and Sr$^{2+}$ ions, resulting in quasi-one-dimensional magnetic chains.

We aligned the SrCo$_{2}$V$_{2}$O$_{8}$ crystal sample to the ($h0l$) scattering plane. In the scattering plane, nuclear Bragg peaks were observed at $ h =$ even  and $ l =$ even  positions, which is consistent with the selection rule of the $I4_{1}cd$ space group. Below 5 K a new set of magnetic peaks show up at $ h + l =$ odd  positions which are shifted from the nuclear Bragg peaks by the ${\bf Q}_{AF} =(1,0,0)$ magnetic wave vector. This situation facilitates the collection and measurements of the separated Bragg peaks of crystalline and magnetic origins.

\begin{figure}[htbp]
\label{fig2}
\includegraphics[width= 80mm]{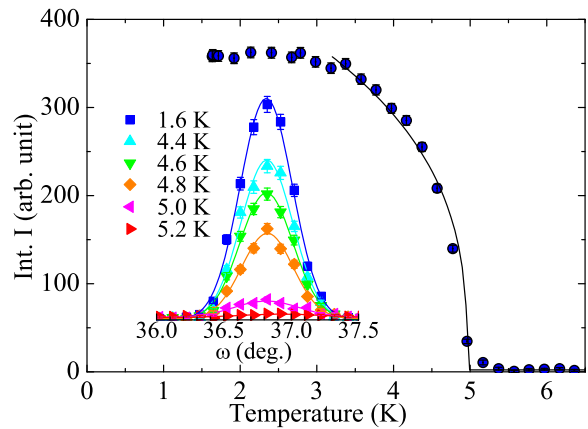}
\caption{(color online). Temperature dependence of the (-2,0,-1) magnetic peak intensity. The solid circles are experimental data, and the solid line is a fit of the experimental data by the function described in the main text. Inset: Rocking scans of (-2,0,-1) Bragg reflection, with increasing temperatures, at 1.6, 4.4, 4.6, 4.8, 5.0 and 5.2 K, respectively.
}
\end{figure}

To investigate the magnetic phase transitions, we measured the temperature dependence of magnetic Bragg reflection $(-2,0,-1)$ from 1.6 K to 8.4 K. It clearly illustrates the emergence of new peaks below the magnetic transition. The integrated intensities are fitted to the function $ I = I_{0} + A(1 - \frac{T}{T_{N}})^{2\beta} $ to obtain the  critical temperature $T_N = 4.96(1)$ K. It is consistent with the $\sim$5 K magnetic transition reported in bulk measurement \cite{He2007667,Lejay2011128}. The rocking scan in the inset of Fig.~2\ref{fig2} has resolution limited peak width and shows no change with temperature. It indicates the long-range nature of the magnetic ordering. 

In addition to the 5 K transition, He {\it et al.}~reported a second anomaly at $\sim 3$ K which is attributed to a transition to a canted antiferromagnetic state in their dynamic magnetic susceptibility work \cite{He2007667}. On the other hand, Lejay {\it et al.}\, found no such transition  \cite{Lejay2011128}. To check the existence of the possible additional transition, we measured a selected set of structural and magnetic Bragg reflections in more than three quadrants of the scattering plane at 1.6 K and 4 K. The integrated intensities of the nuclear reflections at 1.6 K and 4 K don't yield any observable ferromagnetic signal. There is no anomaly either on antiferromagnetic reflections around 3 K. Therefore, our data do not detect the additional transition.

\begin{figure}[tb]
\label{fig3}
\begin{center}
\includegraphics[width=80mm]{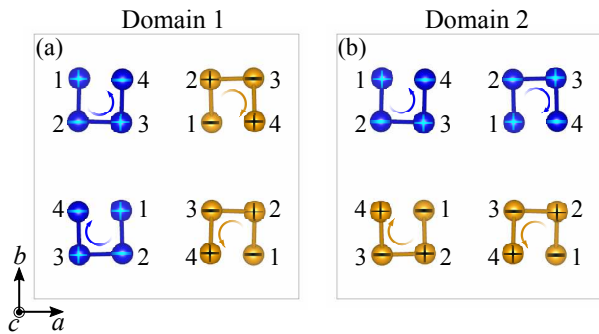}
\caption{(color online). Two magnetic domains of SrCo$_{2}$V$_{2}$O$_{8}$ viewed in $c$ axis. Only intrachain bonded Co atoms are presented, with their sense of rotation marked by screw arrows. The $c$-axis fractional coordinates are $z$, $z$+1/4, $z$+1/2, $z$+3/4 respectively for Co with '1', '2', '3', '4' labels. Spin directions parallel or antiparallel to the $c$ axis are marked by '+' and '-' signs. The same (blue and yellow) colors present ferromagnetic arrangement in basal plane. 
}
\end{center}
\end{figure}

The propagation vector of the magnetic structure ${\bf Q}_{AF} =(1,0,0)$ of SrCo$_{2}$V$_{2}$O$_{8}$ is the same as that of BaCo$_{2}$V$_{2}$O$_{8}$. In BaCo$_{2}$V$_{2}$O$_{8}$, the spins of Co$^{2+}$ are aligned antiferromagnetically along the screw chains, but at a give basal plane they form antiferromagnetic (or ferromagnetic) alignment along the $a$ (or $b$) axis \cite{PhysRevB.87.054408}. 
This spin configuration is no longer invariant under the symmetry operator $4_1$ screw axis of the paramagnetic space group $I4_{1}cd$. Two magnetic domains are thus expected with antiferromagnetic lines formed along either $a$ or $b$ axis, as illustrated in Fig.~3\ref{fig3} (a) and (b) respectively.

The magnetic neutron scattering cross section \cite{9781139107808} of SrCo$_{2}$V$_{2}$O$_{8}$ is described by
\begin{equation}\label{cross section}
\sigma (\textbf{q}) = (\frac{\gamma r_{0}}{2})^{2} N_{m} \dfrac{(2 \pi)^{3}}{v_{m}} \langle M \rangle ^{2} \vert f(q)\vert ^{2} \vert \mathcal{F}\vert ^{2} \langle 1 - (\hat{\textbf{q}} \cdot \hat{\textbf{s}})^{2} \rangle ,
\end{equation}
where $ (\frac{\gamma r_{0}}{2})^{2} = $ 0.07265 $ barn/\mu_{B}^{2} $, $N_{m}$ is the number of magnetic unit cells in the sample, $v_{m}$ is the volume of magnetic unit cell, $M$ is the staggered moment of the  Co$^{2+}$ ion, $f(q)$ is the magnetic form factor of the Co$^{2+}$, $ \hat{\textbf{q}} $ the unit vector of $\textbf{q}$, $ \hat{\textbf{s}} $ the unit vector of the staggered moment, and $\mathcal{F}$ the magnetic structure factor per unit cell. 
Using (x,y,z) as the fractional coordination of the Co atom in the unit cell, and ($hkl$) Miller indices of neutron scattering vector $\textbf{q}$, 
 the squared magnetic structure factor $\vert \mathcal{F}({\bf q}) \vert ^{2}$  of the antiferromagnetic model shown in Fig.~3\ref{fig3}(a) is
\begin{equation}\label{h = 2n + 1, l = 4n}
\vert \mathcal{F}({\bf q})\vert ^{2} = 64  \{\sin (2\pi hx) - \sin (2\pi hy)\}^{2}
\end{equation}
for $h = 2n + 1, l = 4n$, and 
\begin{equation}\label{h = 2n + 1, l = 4n + 2}
\vert \mathcal{F}({\bf q})\vert ^{2} = 64 \{\sin (2\pi hx) + \sin (2\pi hy)\}^{2}
\end{equation}
for $h = 2n + 1, l = 4n + 2$. Namely, for domain 1,
magnetic signals only appear on positions of $h$ = odd and $l$ = even in the ($h0l$) scattering plane.
For domain 2 in Fig.~3\ref{fig3}(b), the magnetic peaks only present when $h$ = even and $l$ = odd, and
\begin{equation}
\label{h = 2n, l = 2n + 1}
\vert \mathcal{F}({\bf q})\vert ^{2} = 64 \{\sin ^{2} (2\pi hx) + \sin ^{2} (2\pi hy)\}
\end{equation}
for $h = 2n, l = 2n + 1$.
 
\begin{figure}[thbp]
\label{fig4}
\includegraphics[width=80mm]{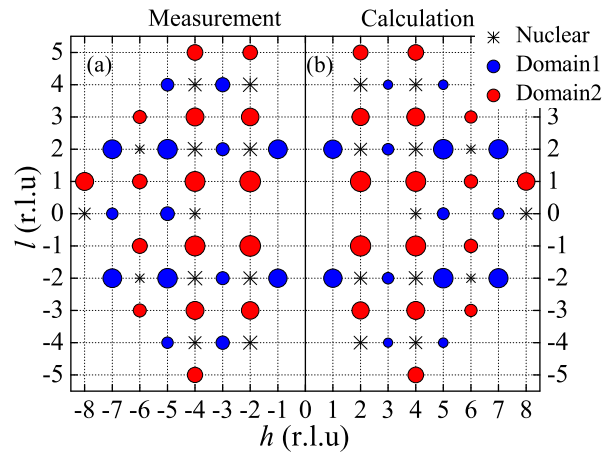}
\caption{(color online). Scattering pattern in reciprocal lattice of ($h0l$) plane. (a) Measured Bragg peaks at 1.6 K. (b) Calculated results. Asterisks mark the structural Bragg peaks. The filled blue circles are magnetic peaks of domain 1, and filled red circles are magnetic peaks of domain 2. The logarithmic scale is used for the magnetic intensities that are illustrated by the size of symbols.
}
\end{figure}

The measured magnetic cross-section deduced from the integrated intensity of the rocking scans by $\sigma (\textbf{q})= I \sin(2\theta)$ and normalized by nuclear Bragg peaks is shown in Fig.~4\ref{fig4}(a).
The blue and red symbols denote Bragg intensities from the domain 1 and 2 shown in Fig.~3, respectively.
The experimental data agree well with the intensities calculated with the same magnetic structure of BaCo$_{2}$V$_{2}$O$_{8}$ \cite{PhysRevB.87.054408} but of a staggered magnetic moment $2.16(1) \mu_{B}$ per Co, see Fig.~4\ref{fig4}(b).
The relative intensities of the blue and red Bragg peaks lead to the domain population $49.9(1)\%$ for domain 1 and $50.1(1)\%$ for domain 2 in the least squared fit. Therefore, the two domains have essentially the equal population.
After we finished our work, we noticed a neutron powder diffraction work on magnetic structure of SrCo$_{2}$V$_{2}$O$_{8}$ \cite{PhysRevB.89.094402}. Both $T_N=5.21(3)$ and $M=2.25(5) \mu_B$/Co are slightly larger than our values. While the same magnetic structure was reported by Bera et al., 
the domain population can be measured only in a single-crystal diffraction experiment and the domain information is important
in achieving the realistic refinement result, as pointed out by Can\'{e}vet {\it et al.}\ \cite{PhysRevB.87.054408}.

The magnetic structure of SrCo$_{2}$V$_{2}$O$_{8}$ is identical to that of the isostructural BaCo$_{2}$V$_{2}$O$_{8}$. Only the Co moment $2.16(1) \mu_{B}$ is slightly smaller than $2.267(3) \mu_{B}$ reported for BaCo$_{2}$V$_{2}$O$_{8}$ \cite{PhysRevB.87.054408}. Our result thus
explains the similar macroscopic properties observed in bulk measurements:
Both SrCo$_{2}$V$_{2}$O$_{8}$ and BaCo$_{2}$V$_{2}$O$_{8}$ show $\lambda$-shape anomalies in heat capacity measurements at  $T_N\sim 5 K$ at zero magnetic field. The antiferromagnetic
order is suppressed by magnetic field in similar manners for both compounds, with the critical field smaller for H $\parallel c$ than H $\perp c$ \cite{He2006458,he2005crystal,Lejay2011128,PhysRevB.90.104419}.
The same in-plain anisotropy between H $\parallel (110)$ and $(100)$ was also recently revealed  \cite{JPSJ.82.033706,PhysRevB.87.224413,PhysRevB.90.104419}.
In the substitution study, all the key features are preserved in Ba$_{1-x}$Sr$_{x}$Co$_{2}$V$_{2}$O$_{8}$ in thermal expansion and magnetostriction measurements \cite{PhysRevB.90.104419}.

As a realization of the one-dimensional XXZ half-integer spin chain model, BaCo$_{2}$V$_{2}$O$_{8}$ has shown
the predicted magnetic field-induced quantum spin liquid behavior \cite{kimura2008novel,PhysRevB.76.224411,PhysRevLett.101.207201,PhysRevB.83.064421,PhysRevB.87.054408}.
Although such field induced phenomena has not been confirmed in SrCo$_{2}$V$_{2}$O$_{8}$,
the same zero-field magnetic state suggests that SrCo$_{2}$V$_{2}$O$_{8}$ could be another interesting
material to explore the novel quantum spin properties.
Recent terahertz spectroscopy experiments show that SrCo$_{2}$V$_{2}$O$_{8}$ is less anisotropic than
BaCo$_{2}$V$_{2}$O$_{8}$ with the anisotropy parameter $\epsilon \approx 0.73$ comparing to $\epsilon \approx 0.46$ for the latter, with $\epsilon=1$ corresponding to the isotropic Heisenberg limit  \cite{PhysRevB.91.140404,kimura2006high}.
Therefore, SrCo$_{2}$V$_{2}$O$_{8}$ from the transition metal vanadates series $AM_{2}$V$_{2}$O$_{8}$ provides an interesting tuning parameter in the anisotropy parameter. Additional tuning parameter, such as that in the spin ladder compounds \cite{PhysRevLett.101.137207, PhysRevLett.101.247202, PhysRevLett.111.106404, PhysRevLett.111.137205}, would enrich study on quasi-one-dimensional quantum antiferromagnetism.

In conclusion, the magnetic structure of SrCo$_{2}$V$_{2}$O$_{8}$ is determined in our neutron single-crystal
diffraction study as the same as that of BaCo$_{2}$V$_{2}$O$_{8}$ \cite{PhysRevB.83.064421}. The magnetic moment of the Co$^{2+}$ ions is  2.16(1) $\mu_{B}$/Co at 1.6 K and the N\'{e}el temperature is 4.96(1) K. The antiferromagnetic structure breaks the four-fold in-plane symmetry and the resulting two magnetic domains have the identical population.
We do not observe the 3 K canting transition, thus SrCo$_{2}$V$_{2}$O$_{8}$ and BaCo$_{2}$V$_{2}$O$_{8}$ share the
same zero field magnetic state. With almost identical magnetic state but less magnetic anisotropy, SrCo$_{2}$V$_{2}$O$_{8}$ could be a worthy addition to BaCo$_{2}$V$_{2}$O$_{8}$ in investigation on spin-chain quantum antiferromagnetism.

The work at RUC and FIRSM was supported by National Basic Research Program of China (Grant Nos.~2012CB921700 and 2011CBA00112) and the National Natural Science Foundation of China (Grant Nos.~11034012 and 11190024).


%

\end{document}